\documentclass[12pt]{article}
\usepackage{amsmath,amssymb,amsthm,amsxtra,overpic,bbm,bm,epsfig,subfigure}
\usepackage{hyperref}
\usepackage{graphicx}
\usepackage{color}
\usepackage{comment}
\usepackage{epstopdf}
\usepackage{float}
\usepackage{cite}
\usepackage{overpic} 
\usepackage{hyperref}
\usepackage{slashed,stmaryrd}

\usepackage{lscape}
\usepackage{array}
\usepackage{booktabs}

\def\thefootnote{\fnsymbol{footnote}}

\begin{document}
\vspace{0.2cm}

\begin{center}
{\Large\bf On the Properties of the Effective Jarlskog Invariant for Three-flavor Neutrino Oscillations in Matter}
\end{center}

\vspace{0.2cm}

\begin{center}
{\bf Xin Wang}~$^{a,~b,~}$\footnote{E-mail: wangx@ihep.ac.cn},
\quad
{\bf Shun Zhou}~$^{a,~b,~}$\footnote{E-mail: zhoush@ihep.ac.cn}
\\
{\small $^a$Institute of High Energy Physics, Chinese Academy of
Sciences, Beijing 100049, China \\
$^b$School of Physical Sciences, University of Chinese Academy of Sciences, Beijing 100049, China}
\end{center}

\vspace{1.5cm}

\begin{abstract}
In this paper, we show that the ratio of the effective Jarlskog invariant $\widetilde{\cal J}$ for leptonic CP violation in three-flavor neutrino oscillations in matter to its counterpart ${\cal J}$ in vacuum $\widetilde{\cal J}/{\cal J} \approx 1/(\widehat{C}^{}_{12} \widehat{C}^{}_{13})$ holds as an excellent approximation, where $\widehat{C}^{}_{12} \equiv \sqrt{1 - 2 \widehat{A}^{}_* \cos 2\theta^{}_{12} + \widehat{A}^2_*}$ with $\widehat{A}^{}_* \equiv a\cos^2 \theta^{}_{13}/\Delta^{}_{21}$ and $\widehat{C}^{}_{13} \equiv \sqrt{1 - 2 A^{}_{\rm c} \cos 2\theta^{}_{13} + A^2_{\rm c}}$ with $A^{}_{\rm c} \equiv a/\Delta^{}_{\rm c}$. Here $\Delta^{}_{ij} \equiv m^2_i - m^2_j$ (for $ij = 21, 31, 32$) stand for the neutrino mass-squared differences in vacuum and $\theta^{}_{ij}$ (for $ij = 12, 13, 23$) are the neutrino mixing angles in vacuum, while $\Delta^{}_{\rm c} \equiv \Delta^{}_{31}\cos^2\theta^{}_{12} + \Delta^{}_{32} \sin^2 \theta^{}_{12}$  and the matter parameter $a \equiv 2\sqrt{2}G^{}_{\rm F} N^{}_e E$ are defined. This result has been explicitly derived by improving the previous analytical solutions to the renormalization-group equations of effective neutrino masses and mixing parameters in matter. Furthermore, as a practical application, such a simple analytical formula has been implemented to understand the existence and location of the extrema of $\widetilde{\cal J}$.
\end{abstract}


\def\thefootnote{\arabic{footnote}}
\setcounter{footnote}{0}

\newpage
\section{Introduction}
The matter effects on three-flavor neutrino oscillations in a medium~\cite{Wolfenstein:1977ue, Mikheev:1986gs} play a very important role in our understanding of various neutrino oscillation experiments~\cite{Kuo:1989qe, Mikheyev:1989dy}. Recently, a complete set of differential equations of the effective neutrino masses $\widetilde{m}^{}_i$ and the effective neutrino mixing matrix elements $V^{}_{\alpha i}$ (for $i = 1, 2, 3$ and $\alpha = e, \mu, \tau$) in ordinary matter with respect to matter parameter $a \equiv 2\sqrt{2}G^{}_{\rm F} N^{}_e E$, where $E$ is the neutrino beam energy, $G^{}_{\rm F}$ is the Fermi constant and $N^{}_e$ is the net electron number density, have been derived in Refs.~\cite{Chiu:2017ckv, Xing:2018} to describe the connection between the fundamental neutrino oscillation parameters in vacuum and those effective ones in matter. In particular, a close analogy of these differential equations with  the renormalization-group equations (RGEs) has been made in Ref.~\cite{Xing:2018}. In the standard parametrization of the effective mixing matrix $V$ in matter~\cite{Tanabashi:2018oca}, the RGEs for three effective mixing angles $\{\widetilde{\theta}^{}_{12}, \widetilde{\theta}^{}_{13}, \widetilde{\theta}^{}_{23}\}$ and the effective CP-violating phase $\widetilde{\delta}$ are found to be~\cite{Xing:2018}
\begin{eqnarray}
\frac{{\rm d}\widetilde{\theta}^{}_{12}}{{\rm d}a}  &=& \dfrac{1}{2} \sin 2\widetilde{\theta}^{}_{12} \left( \widetilde{\Delta}_{21}^{-1} \cos^2 \widetilde{\theta}^{}_{13} - \widetilde{\Delta}_{21}^{} \widetilde{\Delta}_{31}^{-1} \widetilde{\Delta}_{32}^{-1} \sin^2 \widetilde{\theta}^{}_{13} \right) \; , \label{eq:rgetheta12}\\
\frac{{\rm d}\widetilde{\theta}^{}_{13}}{{\rm d}a}  &=& \frac{1}{2} \sin 2\widetilde{\theta}^{}_{13} \left(\widetilde{\Delta}_{31}^{-1} \cos^2 \widetilde{\theta}^{}_{12} + \widetilde{\Delta}_{32}^{-1} \sin^2 \widetilde{\theta}_{12}^{} \right) \; , \label{eq:rgetheta13}\\
\frac{{\rm d}\widetilde{\theta}^{}_{23}}{{\rm d}a}  &=& \frac{1}{2} \widetilde{\Delta}^{}_{21} \widetilde{\Delta}_{31}^{-1} \widetilde{\Delta}_{32}^{-1} \sin 2\widetilde{\theta}^{}_{12} \sin \widetilde{\theta}^{}_{13} \cos \widetilde{\delta} \; , \label{eq:rgetheta23} \\
\frac{{\rm d}\widetilde{\delta}}{{\rm d}a}  &=& - \widetilde{\Delta}^{}_{21} \widetilde{\Delta}_{31}^{-1} \widetilde{\Delta}_{32}^{-1} \sin 2\widetilde{\theta}^{}_{12} \sin \widetilde{\theta}^{}_{13} \sin \widetilde{\delta} \cot 2\widetilde{\theta}^{}_{23} \; ; \label{eq:rgedelta}
\end{eqnarray}
and the RGEs for the effective neutrino mass-squared differences $\widetilde{\Delta}^{}_{ij} \equiv \widetilde{m}^2_i - \widetilde{m}^2_j$ for $ij = 21, 31, 32$ are given by~\cite{Xing:2018}
\begin{eqnarray}
\frac{\mathrm{d} \widetilde{\Delta}_{21}}{\mathrm{d}a} &=& -\cos^2 \widetilde{\theta}^{}_{13} \cos 2\widetilde{\theta}^{}_{12} \; , \label{eq:rgeDel21} \\
\frac{\mathrm{d} \widetilde{\Delta}_{31}}{\mathrm{d}a} &=& \sin^2 \widetilde{\theta}^{}_{13} - \cos^2 \widetilde{\theta}^{}_{13} \cos^2 \widetilde{\theta}^{}_{12} \; , \label{eq:rgeDel31} \\
\frac{\mathrm{d}\widetilde{\Delta}_{32}}{\mathrm{d}a} &=& \sin^2 \widetilde{\theta}^{}_{13} - \cos^2 \widetilde{\theta}^{}_{13} \sin^2 \widetilde{\theta}^{}_{12} \; . \label{eq:rgeDel32}
\end{eqnarray}
These RGEs resemble very much those of leptonic flavor mixing parameters when running from a superhigh-energy scale to the low-energy scale~\cite{Antusch:2003kp, Mei:2005qp, Xing:2005fw, Ohlsson:2013xva}. For instance, the Naumov~\cite{Naumov:1991ju, Krastev:1988yu, Harrison:1999df, Xing:2000ik} and Toshev~\cite{Toshev:1991ku} relations, which were originally derived in the study of matter effects on neutrino oscillations, exist also for the RGE running of fermion masses and flavor mixing parameters in the quark sector and the leptonic sector with Dirac neutrinos~\cite{Xing:2018kto}. 

In our previous work~\cite{Wang:2019yfp}, we have presented for the first time the analytical solutions to those RGEs with some reasonable approximations and obtained compact and simple expressions of all the effective oscillation parameters. For three-flavor neutrino oscillations in matter, the leptonic CP violation can be characterized by the effective Jarlskog invariant $\widetilde{\cal J} \equiv \varepsilon^{}_{\alpha \beta \gamma} \varepsilon^{}_{ijk} {\rm Im}\left[V^{}_{\alpha i} V^*_{\alpha j} V^*_{\beta i} V^{}_{\beta j}\right]$, where $\varepsilon^{}_{\alpha \beta \gamma}$ and $\varepsilon^{}_{ijk}$ are the totally-antisymmetric tensors with $(\alpha, \beta, \gamma)$ and $(i, j, k)$ being the cyclic permutations of $(e, \mu, \tau)$ and $(1, 2, 3)$, respectively. It has been briefly mentioned that the ratio of the matter-corrected Jarlskog invariant $\widetilde{\cal J}$ to the Jarlskog invariant ${\cal J} \equiv \varepsilon^{}_{\alpha \beta \gamma} \varepsilon^{}_{ijk} {\rm Im}\left[U^{}_{\alpha i} U^*_{\alpha j} U^*_{\beta i} U^{}_{\beta j}\right]$ in vacuum~\cite{Jarlskog:1985ht,Wu:1985ea}, where $U$ stands for the leptonic flavor mixing matrix in vacuum, can be approximately expressed as~\cite{Wang:2019yfp}
\begin{eqnarray}\label{eq:Jarl}
\frac{\widetilde{\cal J}}{\cal J} \approx \frac{1}{\sqrt{1 - 2 A^{}_* \cos 2\theta^{}_{12} + A^2_*}} \cdot \frac{1}{\sqrt{1 - 2 A^{}_{\rm c} \cos 2\theta^{}_{13} + A^2_{\rm c}}} \; ,
\end{eqnarray}
where $A^{}_* \equiv a/\Delta^{}_{21}$ and $A^{}_{\rm c} \equiv a/\Delta^{}_{\rm c}$ with $\Delta^{}_{\rm c} \equiv \Delta^{}_{31}\cos^2\theta^{}_{12} + \Delta^{}_{32}\sin^2\theta^{}_{12}$ have been defined~\cite{Minakata:2015gra, Li:2016pzm, Zhou:2016luk}. Here $\Delta^{}_{ij} \equiv m^2_i - m^2_j$ (for $ij = 21, 31, 32$) denote the neutrino mass-squared differences in vacuum and $\theta^{}_{ij}$ (for $ij = 12, 13, 23$) are the neutrino mixing angles in vacuum. Then it has been recognized in Ref.~\cite{Denton:2019yiw} that the simple factorization of the Jarlskog invariant $\widetilde{\cal J}$ in Eq.~(\ref{eq:Jarl}) can be further improved by replacing $A^{}_* = a/\Delta^{}_{21}$ with $\widehat{A}^{}_* \equiv a\cos^2\theta^{}_{13}/\Delta^{}_{21}$. Though in a different context, this replacement has actually been noticed in the treatment of matter effects on the flavor conversions of solar neutrinos in the two-flavor approximation~\cite{Blennow:2003xw, Huang:2018ufu}.  

In the present work, we improve the analytical solutions to the RGEs in Eqs.~(\ref{eq:rgetheta12})-(\ref{eq:rgeDel32}), following the approach suggested in Ref.~\cite{Wang:2019yfp}, and demonstrate that the improvement  can be simply achieved by relaxing the approximation of $\cos^2\theta^{}_{13} \approx 1$. Moreover, we apply the improved formulas to the study of the basic properties of the matter-corrected Jarlskog invariant $\widetilde{\cal J}$. As we shall show later, the extrema of $\widetilde{\cal J}$ can be easily found and explained. Such an investigation is instructive for our understanding of the matter effects on the leptonic CP violation, which is one of the primary goals of future long-baseline accelerator neutrino oscillation experiments~\cite{Branco:2011zb}.

The remaining part of this work is structured as follows. In Sec.~\ref{sec:imp}, we present the improvement on the analytical expressions of the effective neutrino oscillation parameters in matter. Then, the analytical results are implemented to investigate the extrema of the Jarlskog invariant $\widetilde{\cal J}$ in Sec.~\ref{sec:app}. Finally, we summarize our main conclusions in Sec.~\ref{sec:conc}.

\section{Improved Analytical Solutions}\label{sec:imp}

Let us first explain how to improve the analytical solutions to the RGEs in Ref.~\cite{Wang:2019yfp}. For clarity, we focus only on three-flavor neutrino oscillations in matter in the case of normal neutrino mass ordering (NO) with $m^{}_1 < m^{}_2 < m^{}_3$ (or $\Delta^{}_{31} > 0$). The case of inverted neutrino mass ordering (IO) with $m^{}_3 < m^{}_1 < m^{}_2$ (or $\Delta^{}_{31} < 0$) can be examined in a similar way. For antineutrino oscillations, one can repeat the same calculations after the replacements $U \to U^*$ and $a\to -a$.

The starting point to analytically solve the RGEs in Eqs.~(\ref{eq:rgetheta12})-(\ref{eq:rgeDel32}) is to series expand $\widetilde{\Delta}^{}_{ij}$ (for $ij = 21, 31, 32$) in terms of the perturbation parameter $\alpha \equiv \Delta^{}_{21}/\Delta^{}_{31}$, which is estimated as $\alpha \approx 0.03$ for $\Delta^{}_{21} \approx 7.39\times 10^{-5}~{\rm eV}^2$ and $\Delta^{}_{31} \approx 2.523\times 10^{-3}~{\rm eV}^2$,  as first done in Ref.~\cite{Freund:2001}. It has been shown in Ref.~\cite{Wang:2019yfp} that the series expansions of $\widetilde{\Delta}^{}_{ij}$ will be much simpler if the perturbation parameter is chosen as $\alpha^{}_{\rm c} \equiv \Delta^{}_{21}/\Delta^{}_{\rm c}$, which is on the same order of $\alpha$ and thus can be equally good for perturbation calculations. More explicitly, to the first order of $\alpha^{}_{\rm c}$, we have
\begin{eqnarray}
\widetilde{\Delta}^{}_{21} &\approx& \Delta^{}_{\rm c} \left[\left(1 + A^{}_{\rm c} - \widehat{C}^{}_{13}\right)/2  - \alpha^{}_{\rm c} \cos 2\theta^{}_{12} \right] \; , \label{eq:Del21c} \\
\widetilde{\Delta}^{}_{31} &\approx& \Delta^{}_{\rm c} \left[\left(1 + A^{}_{\rm c} + \widehat{C}^{}_{13}\right)/2  - \alpha^{}_{\rm c} \cos 2\theta^{}_{12} \right] \; , \label{eq:Del31c} \\
\widetilde{\Delta}^{}_{32} &\approx& \Delta^{}_{\rm c} \widehat{C}^{}_{13} \; , \label{eq:Del32c}
\end{eqnarray}
with $\widehat{C}^{}_{13} \equiv \sqrt{1 - 2 A^{}_{\rm c} \cos 2\theta^{}_{13} + A^2_{\rm c}}$. One can observe that the first-order term ${\cal O}(\alpha^{}_{\rm c})$ on the right-hand side of Eq.~(\ref{eq:Del32c}) is absent. The series expansions in Eqs.~(\ref{eq:Del21c})-(\ref{eq:Del32c}) can be regarded as the trial solutions to $\widetilde{\Delta}^{}_{ij}$. Inserting these solutions into their RGEs in Eqs. (\ref{eq:rgeDel21})-(\ref{eq:rgeDel32}), one can obtain the analytical solution to $\widetilde{\theta}^{}_{13}$ as follows
\begin{eqnarray}
\cos^2 \widetilde{\theta}^{}_{13} = \frac{1}{2} \left( 1 - \frac{A^{}_{\rm c} - \cos 2\theta^{}_{13}}{\widehat{C}^{}_{13}} \right) \; .
\label{eq:solth13no}
\end{eqnarray}
However, in order to find out the solution to the mixing angle $\widetilde{\theta} ^{}_{12}$, we have to modify the expressions of $\widetilde{\Delta}^{}_{ij}$ in Eqs.~(\ref{eq:Del21c})-(\ref{eq:Del32c}) in the following way
\begin{eqnarray}
\widetilde{\Delta}^{}_{21} &=& \Delta^{}_{\rm c} \left[(1 + A^{}_{\rm c} - \widehat{C}^{}_{13})/2 + \alpha^{}_{\rm c} ({\cal F} - {\cal G}) \right] \; , \label{eq:Del21n} \\
\widetilde{\Delta}_{31} &=& \Delta^{}_{\rm c} \left[(1 + A^{}_{\rm c} + \widehat{C}^{}_{13})/2 + \alpha^{}_{\rm c} {\cal F} \right] \; , \label{eq:Del31n} \\
\widetilde{\Delta}_{32} &=& \Delta^{}_{\rm c} \left( \widehat{C}^{}_{13} + \alpha^{}_{\rm c} {\cal G} \right) \; , \label{eq:Del32n}
\end{eqnarray}
where ${\cal F}(A^{}_{\rm c})$ and ${\cal G}(A^{}_{\rm c})$ are two functions of $A^{}_{\rm c}$ that need to be determined. The main strategy to determine these two functions is to require that the exact RGEs in Eqs.~(\ref{eq:rgetheta12})-(\ref{eq:rgeDel32}) must be satisfied if we substitute Eqs. (\ref{eq:Del21n})-(\ref{eq:Del32n}) into them. After some straightforward calculations, we finally derive the differential equation of ${\cal F}(A^{}_{\rm c})$, i.e.,
\begin{equation}
\frac{{\rm d}{\cal F}}{{\rm d}A^{}_{\rm c}} = \frac{(A^{}_{\rm c} - \widehat{C}^{}_{13} - 1) (\cos 2\theta^{}_{12} + {\cal F}) \cos^2 \theta^{}_{13} } {\widehat{C}^2_{13} \left[ 2(\cos 2\theta^{}_{12} + 2{\cal F}) \alpha^{}_{\rm c} + (1 + A^{}_{\rm c} - \widehat{C}^{}_{13})\right]} \; ,
\label{eq:dFdA}
\end{equation}
with the initial condition ${\cal F}(0) = \sin^2 \theta^{}_{12}$. In our previous work~\cite{Wang:2019yfp}, the approximate relations $(A^{}_{\rm c} - \widehat{C}^{}_{13} - 1)/\widehat{C}^2_{13} \approx -2$ and $(1 + A^{}_{\rm c} - \widehat{C}^{}_{13})/2 \approx A^{}_{\rm c} \cos^2 \theta^{}_{13}$ in the limit of $A^{}_{\rm c} \rightarrow 0$ have been utilized, and the terms proportional to $A^{}_{\rm c} \sin^2 \theta^{}_{13}$ have been ignored. With these approximations, Eq.~(\ref{eq:dFdA}) will be reduced to Eq.~(26) in Ref.~\cite{Wang:2019yfp}. Now we maintain all the terms proportional to $A^{}_{\rm c} \sin^2 \theta^{}_{13}$, and then Eq.~(\ref{eq:dFdA}) can be rewritten as
\begin{eqnarray}
\frac{{\rm d}{\cal F}}{{\rm d}(A^{}_{\rm c} \cos^{2}_{}\theta^{}_{13})} = - \frac{\cos 2\theta^{}_{12} + {\cal F}}{(\cos 2\theta^{}_{12} + 2{\cal F})\alpha^{}_{\rm c} + A^{}_{\rm c}\cos^{2}_{}\theta^{}_{13}} \; ,
\label{eq:dFdAsim}
\end{eqnarray}
to which the exact solution is
\begin{eqnarray}
{\cal F}(A^{}_{\rm c}) = \frac{1}{2}\left[\sqrt{\left(\frac{A^{}_{\rm c}\cos^{2}_{}\theta^{}_{13}}{\alpha^{}_{\rm c}} -  \cos 2\theta^{}_{12}\right)^{2}_{} + \sin^{2}_{} 2\theta^{}_{12}} - \left(\frac{A^{}_{\rm c}\cos^{2}_{}\theta^{}_{13}}{\alpha^{}_{\rm c}} + \cos 2\theta^{}_{12}\right)\right] \; .
\label{eq:solF}
\end{eqnarray}
Substitute Eq. (\ref{eq:solF}) into Eq. (23) of Ref.~\cite{Wang:2019yfp}, we get the improved solution to $\cos^2 \widetilde{\theta}^{}_{12}$, namely,
\begin{equation}
\cos^2 \widetilde{\theta}^{}_{12} =  \frac{1}{2} \left(1 - \frac{\widehat{A}^{}_* - \cos 2\theta^{}_{12}}{\widehat{C}^{}_{12}}\right) \frac{2\widehat{C}^{}_{13} \cos^2 \theta^{}_{13}}{\widehat{C}^{}_{13} - A^{}_{\rm c} + \cos 2\theta^{}_{13}} \; ,
\label{eq:solth12}
\end{equation}
where $\widehat{A}^{}_{\ast} \equiv A^{}_{\rm c}\cos^{2}_{}\theta^{}_{13}/\alpha^{}_{\rm c} = a \cos^{2}_{}\theta^{}_{13}/\Delta^{}_{21}$ and $\widehat{C}^{}_{12 } \equiv \sqrt{1 - 2\widehat{A}^{}_{\ast}\cos 2\theta^{}_{12} + \widehat{A}^{2}_{\ast}}$ have been defined. Comparing the improved solution in Eq.~(\ref{eq:solth12}) with the original one in Ref.~\cite{Wang:2019yfp}, one can immediately realize that the only change is the replacement of $A^{}_{\ast} \equiv a /\Delta^{}_{21}$ by $\widehat{A}^{}_{\ast} \equiv a \cos^{2}_{}\theta^{}_{13}/\Delta^{}_{21}$. 

\begin{figure}[t!]
	\centering
	\includegraphics[width=0.6\textwidth]{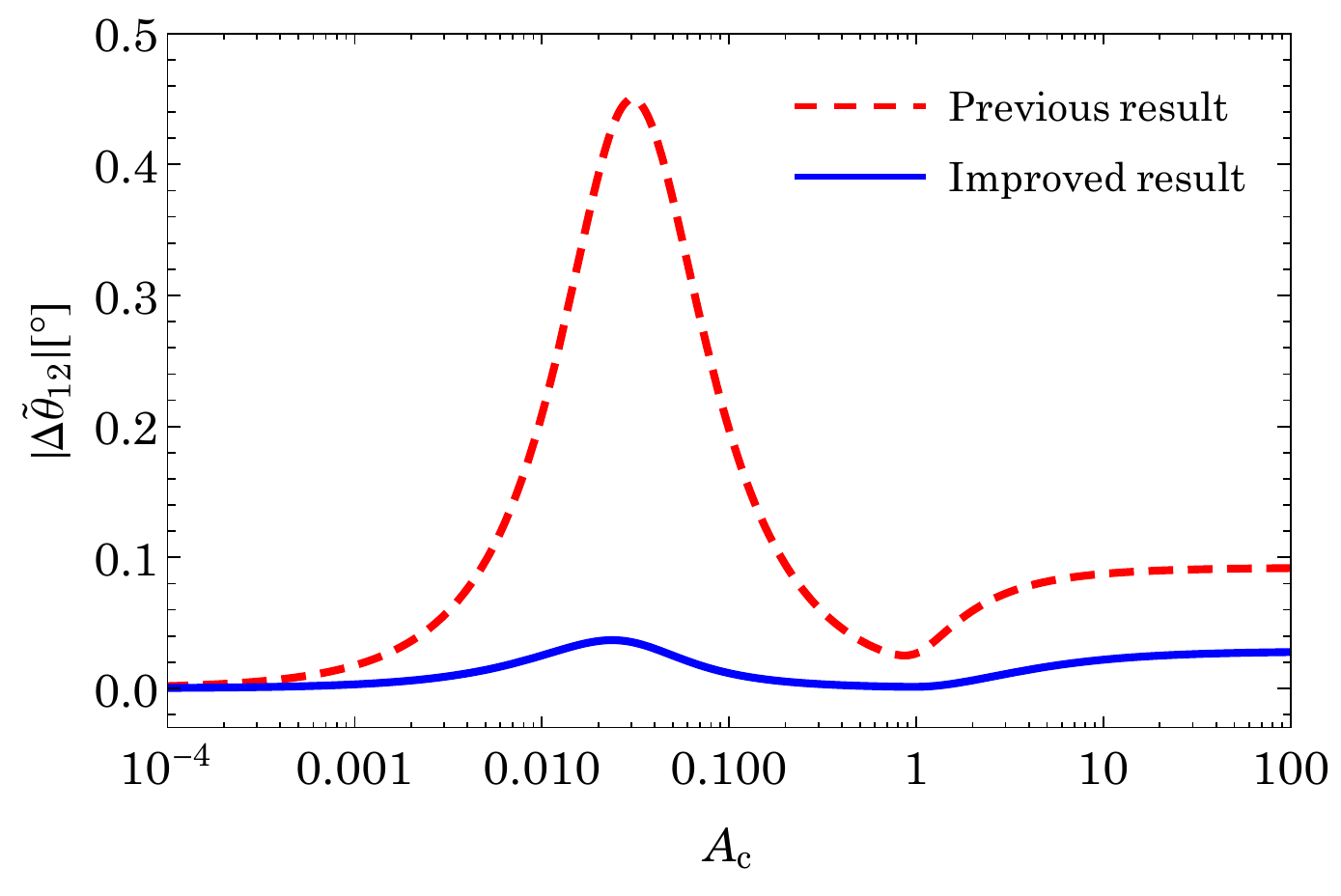}
\vspace{-0.0cm}
	\caption{The absolute value of the difference $\Delta \widetilde{\theta}^{}_{12} \equiv \widetilde{\theta}^{}_{12}|_{\rm analytical} - \widetilde{\theta}^{}_{12}|_{\rm numerical}$ between the approximate analytical result and the exact numerical result, where the red dashed curve corresponds to the previous analytical result while the blue solid curve refers to the improved one in this work.}
	\label{fig:th12} 
\end{figure}

To illustrate how much the improvement on the solution to $\widetilde{\theta}^{}_{12}$ is, we calculate the difference $\Delta \widetilde{\theta}^{}_{12} \equiv \widetilde{\theta}^{}_{12}|_{\rm analytical} - \widetilde{\theta}^{}_{12}|_{\rm numerical}$ between the approximate analytical result and the exact numerical one and show its absolute value $|\Delta \widetilde{\theta}^{}_{12}|$ in Fig.~\ref{fig:th12}, where the best-fit values of neutrino oscillation parameters $\sin^2 \theta^{}_{12} = 0.310$, $\sin^2 \theta^{}_{13} = 0.02241$, $\Delta^{}_{21} = 7.39\times 10^{-5}~{\rm eV}^2$ and $\Delta^{}_{31} = 2.523\times 10^{-3}~{\rm eV}^2$ from the latest global-fit analysis of neutrino oscillation data have been used~\cite{Esteban:2018azc}. The red dashed curve refers to the analytical formula in Ref.~\cite{Wang:2019yfp}, while the blue solid one denotes the result in Eq.~(\ref{eq:solth12}). Although the difference between the previous approximate result and the exact value of $\widetilde{\theta}^{}_{12}$ is always below $0.5^\circ$, the improvement represented by the blue solid curve is quite remarkable. In the latter case, the deviation from the exact value of $\widetilde{\theta}^{}_{12}$ appears to be maximal in the resonance region, which turns out to be at most $0.03^\circ$. It is interesting to notice that the improvement on the analytical result can be attributed to the factor $\cos^2 \theta^{}_{13}$ in the matter term, which has already been observed in the treatment of matter effects on solar neutrino flavor conversions in the two-flavor approximation~\cite{Blennow:2003xw,Huang:2018ufu}. Based on the above observations, we can conclude that the effective mixing angles $\widetilde{\theta}^{}_{12}$ and $\widetilde{\theta}^{}_{13}$ are excellently described by the two-flavor neutrino oscillations in matter~\cite{Wang:2019yfp, Denton:2016wmg, Ioannisian:2018qwl, Xing:2019owb}. In the former case, the neutrino mass-squared difference in vacuum is $\Delta^{}_{21}$, the mixing angle in vacuum is $\theta^{}_{12}$, and the matter parameter is $a \cos^2\theta^{}_{13}$. In the latter case, these three parameters are given by $\Delta^{}_{\rm c}$, $\theta^{}_{13}$ and $a$, respectively.
Given the improved result of the function ${\cal F}$ in Eq.~(\ref{eq:solF}), one can verify that the expressions of three effective neutrino mass-squared differences $\widetilde{\Delta}^{}_{ij}$ (for $ij=21,31,32$) are then given by
\begin{eqnarray}
\widetilde{\Delta}^{}_{21} &=& \Delta^{}_{\rm c} \left[\frac{1}{2}(1 + A^{}_{\rm c} - \widehat{C}^{}_{13}) + (\widehat{C}^{}_{12} - \widehat{A}^{}_*) \alpha^{}_{\rm c}\right] \; , \label{eq:Del21rev} \\
\widetilde{\Delta}^{}_{31} &=& \Delta^{}_{\rm c} \left[\frac{1}{2}(1 + A^{}_{\rm c} + \widehat{C}^{}_{13}) + \frac{1}{2} (\widehat{C}^{}_{12} - \widehat{A}^{}_* - \cos 2\theta_{12}) \alpha^{}_{\rm c} \right] \; , \label{eq:Del31rev} \\ \widetilde{\Delta}^{}_{32} &=& \Delta^{}_{\rm c} \left[\widehat{C}^{}_{13} + \frac{1}{2} (\widehat{A}^{}_* - \widehat{C}^{}_{12} - \cos 2\theta_{12}) \alpha^{}_{\rm c} \right] \; , \label{eq:Del32rev}
\end{eqnarray}
where $\widehat{C}^{}_{12} = \sqrt{1 - 2\widehat{A}^{}_{\ast}\cos 2\theta^{}_{12} + \widehat{A}^{2}_{\ast}}$ with $\widehat{A}^{}_{\ast} \equiv a \cos^{2}_{}\theta^{}_{13}/\Delta^{}_{21}$ should be noted. Since $\widetilde{\theta}^{}_{23} \approx \theta^{}_{23}$ and $\widetilde{\delta} \approx \delta$ are justified, as shown in Ref.~\cite{Wang:2019yfp}, we now get a complete set of approximate analytical solutions to all the effective mass-squared differences, mixing angles and the CP-violating phase.

As an immediate application, the effective Jarlskog invariant $\widetilde{\cal J}$ and all the elements of the effective flavor mixing matrix $|V^{}_{\alpha i}|^2$ (for $\alpha = e, \mu, \tau$ and $i = 1, 2, 3$) can be explicitly calculated.
\begin{itemize}
\item {\it The effective Jarlskog invariant in matter} -- In the standard parametrization of the effective flavor mixing matrix $V$, the effective Jarlskog invariant turns out to be
\begin{equation}
\widetilde{\mathcal{J}} = \sin \widetilde{\theta}^{}_{12} \cos \widetilde{\theta}^{}_{12} \sin \widetilde{\theta}^{}_{13} \cos^2 \widetilde{\theta}^{}_{13} \sin \widetilde{\theta}^{}_{23} \cos \widetilde{\theta}^{}_{23} \sin \widetilde{\delta} \; .
\label{eq:jarls}
\end{equation}
On the other hand, it is well known that the Toshev relation $
\sin 2\widetilde{\theta}^{}_{23} \sin \widetilde{\delta} = \sin 2\theta^{}_{23} \sin \delta$ holds exactly~\cite{Toshev:1991ku}. Adopting the standard parametrization for the flavor mixing matrix $U$ in vacuum, one can also express the Jarlskog invariant ${\cal J}$ in vacuum in terms of the fundamental mixing angles $(\theta^{}_{12}, \theta^{}_{13}, \theta^{}_{23})$ and the CP-violating phase $\delta$, similar to that for the effective Jarlskog invariant $\widetilde{\cal J}$ in Eq.~(\ref{eq:jarls}). Therefore, it is straightforward to get
\begin{equation}
\frac{\widetilde{\mathcal{J}}}{\mathcal{J}} = \frac{\sin 2\widetilde{\theta}^{}_{12} \sin 2\widetilde{\theta}^{}_{13}\cos \widetilde {\theta}^{}_{13}}{\sin 2\theta^{}_{12} \sin 2\theta^{}_{13}\cos \theta^{}_{13}}  \; ,
\label{eq:jratio}
\end{equation}
where the Toshev relation has been used. Now that the analytical formulas of $\cos^2 \widetilde{\theta}^{}_{13}$ and $\cos^2 \widetilde{\theta}^{}_{12}$ are given in Eq.~(\ref{eq:solth13no}) and Eq.~(\ref{eq:solth12}), respectively, we can find 
\begin{eqnarray}
\sin 2\widetilde{\theta}^{}_{13} \approx 2\sqrt{\frac{1}{2}\left(1 - \frac{A^{}_{\rm c} - \cos 2\theta^{}_{13}}{\widehat{C}^{}_{13}}\right)}\cdot \sqrt{\frac{1}{2}\left(1 + \frac{A^{}_{\rm c} - \cos 2\theta^{}_{13}}{\widehat{C}^{}_{13}}\right)} = \frac{\sin 2\theta^{}_{13}}{\widehat{C}^{}_{13}}\; ,
\label{eq:sin2t13}
\end{eqnarray}
and
\begin{eqnarray}
\sin 2\widetilde{\theta}^{}_{12} &\approx& 2\sqrt{\frac{1}{2}\left(1 - \frac{\widehat{A}^{}_* - \cos 2\theta^{}_{12}}{\widehat{C}^{}_{12}}\right)\frac{\cos^2 \theta^{}_{13}}{\cos^2 \widetilde{\theta}^{}_{13}}} \cdot \sqrt{1 - \frac{1}{2}\left(1 - \frac{\widehat{A}^{}_* - \cos 2\theta^{}_{12}}{\widehat{C}^{}_{12}}\right)\frac{\cos^2 \theta^{}_{13}}{\cos^2 \widetilde{\theta}^{}_{13}}} \nonumber \\
&\approx& 2 \sqrt{\frac{1}{2}\left(1 - \frac{\widehat{A}^{}_* - \cos 2\theta^{}_{12}}{\widehat{C}^{}_{12}}\right)\frac{\cos^2 \theta^{}_{13}}{\cos^2 \widetilde{\theta}^{}_{13}}} \cdot \sqrt{1 - \frac{1}{2}\left(1 - \frac{\widehat{A}^{}_* - \cos 2\theta^{}_{12}}{\widehat{C}^{}_{12}}\right)} \nonumber \\
&=& \frac{\cos\theta^{}_{13}}{\cos\widetilde{\theta}^{}_{13}}\frac{\sin 2\theta^{}_{12}}{\widehat{C}^{}_{12}} \; ,
\label{eq:sin2t12}
\end{eqnarray}
where the approximation $\sin^2 \widetilde{\theta}^{}_{12} \approx \left[1 + (\widehat{A}^{}_* - \cos 2\theta^{}_{12})/\widehat{C}^{}_{12}\right]/2$ has been made in the second line of Eq.~(\ref{eq:sin2t12}). Such an approximation is reasonable in the whole range of $A^{}_{\rm c}$. For small values of $A^{}_{\rm c}$, the matter effect is negligible for $\widetilde{\theta}^{}_{13}$ such that $\cos^2 \widetilde{\theta}^{}_{13} \approx \cos^2 \theta^{}_{13}$. In this case, we have $\sin 2\widetilde{\theta}^{}_{12} = \sin 2\theta^{}_{12}/\widehat{C}^{}_{12}$ as in the scenario of two-flavor neutrino oscillations. For large values of $A^{}_{\rm c}$, the mixing angle $\widetilde{\theta}^{}_{13}$ is significantly enhanced, and the approximation of $\cos^2 \widetilde{\theta}^{}_{13} \approx \cos^2 \theta^{}_{13}$ is no longer valid. However, for such a large value of $A^{}_{\rm c}$, the mixing angle $\widetilde{\theta}^{}_{12}$ has already been enhanced to be close to $90^\circ$, namely, $[1 - (\widehat{A}^{}_* - \cos 2\theta^{}_{12})/\widehat{C}^{}_{12}]/2$ approaches zero. Thus, it is safe to ignore the factor of $\cos^2\theta^{}_{13}/\cos^2 \widetilde{\theta}^{}_{13}$ in the second square root in the second line of Eq.~(\ref{eq:sin2t12}). 
Substituting Eqs.~(\ref{eq:sin2t13}) and (\ref{eq:sin2t12}) into Eq.~(\ref{eq:jratio}), we obtain
\begin{eqnarray}
\frac{\widetilde{\cal J}}{\cal J}=\frac{1}{\widehat{C}^{}_{12} \widehat{C}^{}_{13}} \;,
\label{eq:jexp}
\end{eqnarray}
which takes the same form as in Eq.~(\ref{eq:Jarl}) but now with $\widehat{C}^{}_{12} = \sqrt{1 - 2\widehat{A}^{}_{\ast}\cos 2\theta^{}_{12} + \widehat{A}^{2}_{\ast}}$. In fact, this simple relation can also be understood via the Naumov relation~\cite{Naumov:1991ju, Krastev:1988yu, Harrison:1999df, Xing:2000ik}
\begin{eqnarray}
\frac{\widetilde{\cal J}}{\cal J}=\frac{\Delta^{}_{21} \Delta^{}_{31} \Delta^{}_{32}}{\widetilde{\Delta}^{}_{21} \widetilde{\Delta}^{}_{31} \widetilde{\Delta}^{}_{32}} \; \; .
\label{eq:naumov}
\end{eqnarray}
For $A^{}_{\rm c} \ll \cos 2\theta^{}_{13}$, expanding $\widehat{C}^{}_{13}$ on the right-hand side of Eq.~(\ref{eq:Del21rev}) in terms of $A^{}_{\rm c}$ and keeping only the leading-order term, we have $\widetilde{\Delta}^{}_{21} \approx \Delta^{}_{21}\widehat{C}^{}_{12}$. Hence the ratio $\Delta^{}_{21}/\widetilde{\Delta}^{}_{21}$ on the right-hand side of Eq.~(\ref{eq:naumov}) contributes a factor of $1/\widehat{C}^{}_{12}$. On the other hand, $\widetilde{\Delta}^{}_{31} \approx \Delta^{}_{\rm c}$ and $\widetilde{\Delta}^{}_{32} \approx \Delta^{}_{\rm c}\widehat{C}^{}_{13}$ hold at the leading order, contributing the desired factor $(\Delta^{}_{31}\Delta^{}_{32})/(\widetilde{\Delta}^{}_{31}\widetilde{\Delta}^{}_{32}) \approx 1/\widehat{C}^{}_{13}$. For $A^{}_{\rm c} \gtrsim \cos 2\theta^{}_{13}$, these approximations are not justified. However, in the limit of $A^{}_{\rm c} \to \infty$, one can observe $\Delta^{}_{21}/\widetilde{\Delta}^{}_{21} \to \alpha^{}_{\rm c}$ and  $(\Delta^{}_{31}\Delta^{}_{32})/(\widetilde{\Delta}^{}_{31}\widetilde{\Delta}^{}_{32})  \to 1/A^{2}_{\rm c}$. Hence $\widetilde{\cal J}/{\cal J} \to \alpha^{}_{\rm c}/A^{2}_{\rm c}$ has the same asymptotic behavior as $1/(\widehat{C}^{}_{12}\widehat{C}^{}_{13})$ does. To some extent, this is the reason why the simple relation in Eq.~(\ref{eq:jexp}) is also valid in the region of large $A^{}_{\rm c}$.

\item {\it The effective mixing matrix elements in matter} -- With the approximate analytical results of three effective neutrino mixing angles $\widetilde{\theta}^{}_{ij}$ (for $ij = 12, 13, 23$) and the effective CP-violating phase $\widetilde{\delta}$, we can also explicitly write down the expressions of the moduli for all the elements of the effective flavor mixing matrix $V$ according to its standard parametrization. Although two of them have been given and briefly discussed in Ref.~\cite{Wang:2019yfp}, we collect all the analytical expressions below for completeness. The moduli of the matrix elements in the first row are
\begin{eqnarray}
&~& |V^{}_{e 1}|^{2}_{} = \frac{\cos^{2}_{}\theta^{}_{13}}{2}\left( 1-\frac{\widehat{A}^{}_{\ast}-\cos 2 \theta^{}_{12}}{\widehat{C}^{}_{12}} \right) \; , \nonumber \\
&~& |V^{}_{e 2}|^{2}_{} = \frac{1}{2}\left( 1-\frac{A^{}_{\rm c}-\cos 2 \theta^{}_{13}}{\widehat{C}^{}_{13}} \right)-\frac{\cos^{2}_{}\theta^{}_{13}}{2}\left(1-\frac{\widehat{A}^{}_{\ast}-\cos2\theta^{}_{12}}{\widehat{C}^{}_{12}}\right)\; , \nonumber \\ 
&~& |V^{}_{e 3}|^{2}_{} = \frac{1}{2}\left( 1+\frac{A^{}_{\rm c}-\cos 2 \theta^{}_{13}}{\widehat{C}^{}_{13}} \right) \; ; \label{eq:Ve3}
\end{eqnarray}
and those of the elements in the second row are~\footnote{It is worthwhile to point out that there are two typographical errors in Eq.~(42) of Ref.~\cite{Wang:2019yfp}, where the sign in front of the last term on the right-hand side should be reversed and the factor $\widehat{C}^{}_{13}$ in the numerator of this term should be moved to the denominator.}
\begin{eqnarray}
&~& |V^{}_{\mu 1}|^{2}_{} = \cos^{2}_{}\theta^{}_{23}-\left(\sin^{2}_{}\theta^{}_{23}+\frac{2\cos 2\theta^{}_{23}\widehat{C}^{}_{13}}{\widehat{C}^{}_{13}-A^{}_{\rm c}+\cos 2\theta^{}_{13}}\right)|V^{}_{e1}|^{2}_{}+\frac{4 {\cal J}\cot \delta}{\widehat{C}^{}_{12}(\widehat{C}^{}_{13}-A^{}_{\rm c}+\cos 2\theta^{}_{13})} \; , \nonumber \\
&~& |V^{}_{\mu 2}|^{2}_{} = \cos^{2}_{}\theta^{}_{23}-\left(\sin^{2}_{}\theta^{}_{23}+\frac{2\cos 2\theta^{}_{23}\widehat{C}^{}_{13}}{\widehat{C}^{}_{13}-A^{}_{\rm c}+\cos 2\theta^{}_{13}}\right)|V^{}_{e2}|^{2}_{}-\frac{4 {\cal J}\cot \delta}{\widehat{C}^{}_{12}(\widehat{C}^{}_{13}-A^{}_{\rm c}+\cos 2\theta^{}_{13})} \; , \nonumber \\
&~& |V^{}_{\mu 3}|^{2}_{} = \frac{\sin^{2}_{}\theta^{}_{23}}{2}\left( 1-\frac{A^{}_{\rm c}-\cos 2 \theta^{}_{13}}{\widehat{C}^{}_{13}} \right) \;. \label{eq:Vmu3}
\end{eqnarray}
The normalization condition $|V^{}_{e1}|^2 + |V^{}_{e2}|^2 + |V^{}_{e3}|^2 = 1$ for the first row and that for the second row $|V^{}_{\mu 1}|^2 + |V^{}_{\mu 2}|^2 + |V^{}_{\mu 3}|^2 = 1$ can be easily verified. Although it is obviously possible to express $|V^{}_{\mu 1}|^2$ and $|V^{}_{\mu 2}|^2$ in Eq.~(\ref{eq:Vmu3}) in terms of the fundamental mixing parameters in vacuum and the matter parameter $a$, we have used $|V^{}_{e1}|^2$ and $|V^{}_{e2}|^2$ in Eq.~(\ref{eq:Ve3}) in order to render the formulas to be compact and more readable. Finally, the moduli for the elements in the last row are given by
\begin{eqnarray}
&~& |V^{}_{\tau 1}|^{2}_{} = \sin^{2}_{}\theta^{}_{23}-\left(\cos^{2}_{}\theta^{}_{23}-\frac{2\cos 2\theta^{}_{23}\widehat{C}^{}_{13}}{\widehat{C}^{}_{13}-A^{}_{\rm c}+\cos 2\theta^{}_{13}}\right)|V^{}_{e1}|^{2}_{}-\frac{4 {\cal J}\cot \delta}{\widehat{C}^{}_{12}(\widehat{C}^{}_{13}-A^{}_{\rm c}+\cos 2\theta^{}_{13})} \; , \nonumber \\
&~& |V^{}_{\tau 2}|^{2}_{} = \sin^{2}_{}\theta^{}_{23}-\left(\cos^{2}_{}\theta^{}_{23}-\frac{2\cos 2\theta^{}_{23}\widehat{C}^{}_{13}}{\widehat{C}^{}_{13}-A^{}_{\rm c}+\cos 2\theta^{}_{13}}\right)|V^{}_{e2}|^{2}_{}+\frac{4 {\cal J}\cot \delta}{\widehat{C}^{}_{12}(\widehat{C}^{}_{13}-A^{}_{\rm c}+\cos 2\theta^{}_{13})} \; , \nonumber \\
&~& |V^{}_{\tau 3}|^{2}_{} = \frac{\cos^{2}_{}\theta^{}_{23}}{2}\left( 1-\frac{A^{}_{\rm c}-\cos 2 \theta^{}_{13}}{\widehat{C}^{}_{13}} \right) \; , \label{eq:Vtau3}
\end{eqnarray}
from which one can immediately check that the normalization condition $|V^{}_{\tau 1}|^2 + |V^{}_{\tau 2}|^2 + |V^{}_{\tau 3}|^2 = 1$ is fulfilled. Similarly, we can prove that $|V^{}_{ei}|^2 + |V^{}_{\mu i}|^2 + |V^{}_{\tau i}|^2 = 1$ for $i = 1, 2, 3$ are also satisfied.
\end{itemize}
 
It has been demonstrated in Refs.~\cite{Wang:2019yfp, Denton:2019yiw} and in our previous discussions that the analytical formulas for the effective parameters, or equivalently those for the moduli of the effective mixing matrix elements, agree excellently with the exact numerical results. Therefore, it is definitely interesting to see how they can be applied to the investigation of the leptonic CP violation for three-flavor neutrino oscillations in matter.  

\section{Extrema of the Jarlskog Invariant}\label{sec:app}

\begin{figure}[t!]
	\centering
	\includegraphics[width=\textwidth]{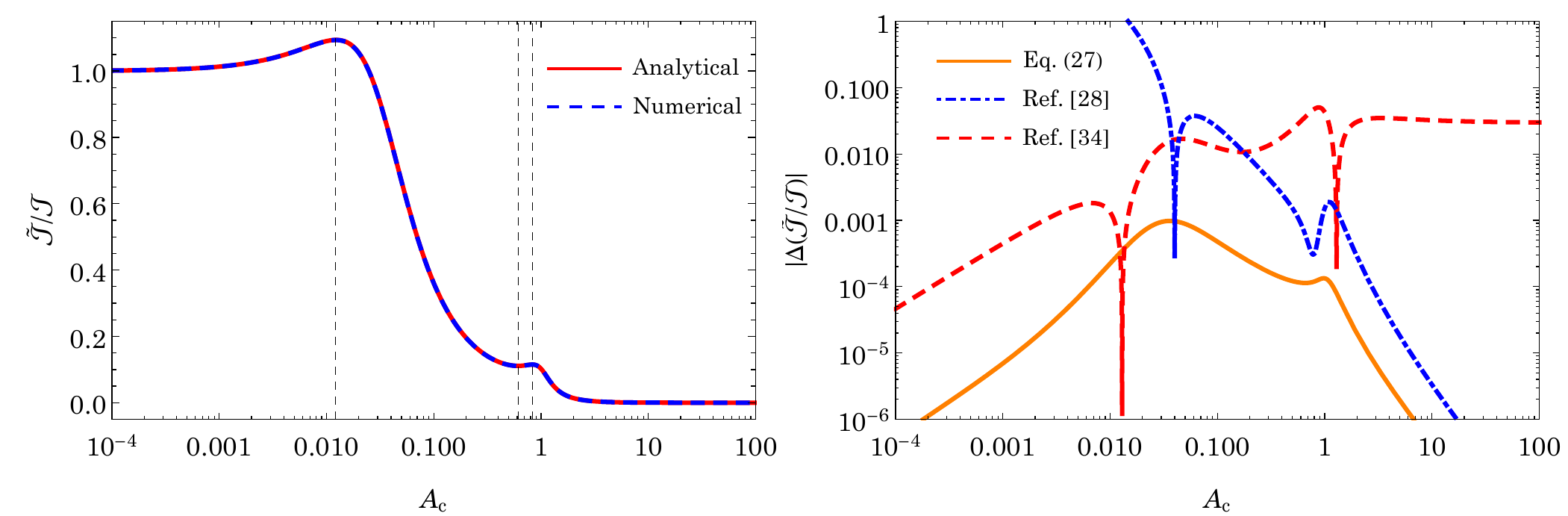}
\vspace{-0.3cm}
	\caption{In the left panel, the analytical result in Eq.~(\ref{eq:jexp})  (the red solid curve) and the exact numerical result (the blue dashed curve) of the ratio $\widetilde{\cal J}/{\cal J}$ in the NO case are plotted, where three vertical dashed lines indicate two local maxima and one local minimum of $\widetilde{\cal J}/{\cal J}$. In the right panel, the absolute value of $\Delta (\widetilde{\cal J}/{\cal J}) \equiv (\widetilde{\cal J}/{\cal J})|^{}_{\rm analytical} - (\widetilde{\cal J}/{\cal J})|^{}_{\rm numerical}$ is presented for three different analytical formulas. The best-fit values of neutrino oscillation parameters from the latest global-fit analysis of neutrino oscillation data have been taken~\cite{Esteban:2018azc}.}
	\label{fig:jarlskog} 
\end{figure}

As we have mentioned, the leptonic CP violation for three-flavor neutrino oscillations in matter can be described by the Jarlskog invariant $\widetilde{\cal J}$~\cite{Krastev:1988yu, Petcov:2018zka}. In the left panel of Fig.~\ref{fig:jarlskog}, we have plotted the ratio $\widetilde{\cal J}/{\cal J}$ by using the analytical formula in Eq.~(\ref{eq:jexp}), where the exact numerical result has also been shown for comparison. An excellent agreement between the analytical (the red solid curve) and numerical (the blue dashed curve) results is evident. In our numerical calculations, the previous best-fit values of neutrino oscillation parameters, together with $\sin^2\theta^{}_{23} = 0.558$ and $\delta = 222^\circ$, from the latest global-fit analysis of neutrino oscillation data have been adopted~\cite{Esteban:2018azc}. In the right panel of Fig.~\ref{fig:jarlskog}, we show the absolute value of the differences between the analytical results and the exact numerical one, where $\Delta (\widetilde{\cal J}/{\cal J}) \equiv (\widetilde{\cal J}/{\cal J})|^{}_{\rm analytical} - (\widetilde{\cal J}/{\cal J})|^{}_{\rm numerical}$ has been defined. The analytical results are taken from Eq.~(\ref{eq:jexp}) in this work and from Refs.~\cite{Freund:2001, Xing:2016ymg}. The accuracy of the analytical formula in Eq.~(\ref{eq:jexp}) is as high as $0.1\%$ for any values of $A^{}_{\rm c}$, which is much better than that of the approximate analytical formulas obtained in Refs.~\cite{Freund:2001, Xing:2016ymg}. 

In this section, we investigate the existence and location of the extrema of the ratio $\widetilde{\cal J}/{\cal J}$, which possesses two local maxima and one local minimum as indicated by three vertical dashed lines in the left panel of Fig.~\ref{fig:jarlskog}. As in the previous section, we focus on the scenario of neutrino oscillations in matter in the NO case. The results for the scenario of antineutrino oscillations and the IO case will be discussed later. As one can observe from Eq.~(\ref{eq:jexp}), the extrema of $\widetilde{\cal J}/{\cal J}$ must be related to the coefficients $\widehat{C}^{}_{12}$ and $\widehat{C}^{}_{13}$,  which are actually the functions of $A^{}_{\rm c}$ and regularize the resonances corresponding to $\Delta^{}_{21}$ and $\Delta^{}_{31}$ for neutrino oscillations in matter in the NO case. To see this point clearly, we have presented in Fig. \ref{fig:C12C13} the evolution of $\widetilde{\cal J}/{\cal J}$, $1/\widehat{C}^{}_{12}$ and $1/\widehat{C}^{}_{13}$ against the matter parameter $A^{}_{\rm c}$, where those two vertical dashed lines denote the resonance of $1/\widehat{C}^{}_{12} = 1/\sqrt{(\widehat{A}^{}_* - \cos 2\theta^{}_{12})^2 + \sin^2 2\theta^{}_{12}}$ at $\widehat{A}^{}_* = \cos 2\theta^{}_{12}$ and that of  $1/\widehat{C}^{}_{13} = 1/\sqrt{(A^{}_{\rm c} - \cos 2\theta^{}_{13})^2 + \sin^2 2\theta^{}_{13}}$ at $A^{}_{\rm c} = \cos 2\theta^{}_{13}$, respectively .

In order to figure out the extrema of $\widetilde{\cal J}/{\cal J}$, we have to compute its first derivative and require it to be vanishing, namely,
\begin{eqnarray}
\frac{{\rm d}}{{\rm d}A^{}_{\rm c}} \left(\frac{\widetilde{\cal J}}{{\cal J}}\right) =  \frac{{\rm d}}{{\rm d}A^{}_{\rm c}} \left(\frac{1}{\widehat{C}^{}_{12} \widehat{C}^{}_{13}}\right) = - \frac{1}{\widehat{C}^2_{12} \widehat{C}^2_{13}} \left[ \widehat{C}^{}_{13} \left(\frac{{\rm d}\widehat{C}^{}_{12}}{{\rm d}A^{}_{\rm c}}\right) + \widehat{C}^{}_{12} \left(\frac{{\rm d}\widehat{C}^{}_{13}}{{\rm d}A^{}_{\rm c}}\right) \right] = 0 \; , \label{eq:firstder}
\end{eqnarray}
where the analytical formula in Eq.~(\ref{eq:jexp}) has been implemented. With the definitions of $\widehat{C}^{}_{12}$ and $\widehat{C}^{}_{13}$, it is quite easy to obtain
\begin{eqnarray}
\frac{{\rm d}\widehat{C}^{}_{13}}{{\rm d}A^{}_{\rm c}} = \frac{1}{\widehat{C}^{}_{13}} \left(A^{}_{\rm c} - \cos 2\theta^{}_{13}\right)\; , \quad 
\frac{{\rm d}\widehat{C}^{}_{12}}{{\rm d}A^{}_{\rm c}} = \frac{1}{\widehat{C}^{}_{12}} \left[A^{}_{\rm c} \left(\frac{\cos^2 \theta^{}_{13}}{\alpha^{}_{\rm c}}\right)^2 - \cos 2\theta^{}_{12} \left(\frac{\cos^2 \theta^{}_{13}}{\alpha^{}_{\rm c}}\right) \right] \; . \label{eq:dCdA}
\end{eqnarray}
After inserting Eq.~(\ref{eq:dCdA}) into Eq.~(\ref{eq:firstder}), one can get the cubic equation of $A^{}_{\rm c}$, i.e.,
\begin{eqnarray}
2 A^3_{\rm c} - 3\left(\cos 2\theta^{}_{13} + \frac{\cos 2\theta^{}_{12}}{\cos^2 \theta^{}_{13}} \alpha^{}_{\rm c} \right) A^2_{\rm c} + \left(1 + 4\frac{\cos 2\theta^{}_{12} \cos 2\theta^{}_{13}}{\cos^2\theta^{}_{13}} \alpha^{}_{\rm c} \right)A^{}_{\rm c} - \frac{ \cos 2\theta^{}_{12}}{\cos^2 \theta^{}_{13}} \alpha^{}_{\rm c} = 0 \; , \label{eq:cubic}
\end{eqnarray}
where the tiny corrections of ${\cal O}(\alpha^2_{\rm c})$ to the coefficients of the last two terms on the left-hand side have been neglected. Some comments on the solutions to Eq.~(\ref{eq:cubic}) are helpful.
\begin{itemize}
\item In the leading-order approximation, where the terms of ${\cal O}(\alpha^{}_{\rm c})$ in Eq.~~(\ref{eq:cubic}) are ignored, the cubic equation will be greatly simplified
\begin{eqnarray}
2A^3_{\rm c} - 3 \cos 2\theta^{}_{13} A^2_{\rm c} + A^{}_{\rm c} = 0 \;, \label{eq:cubicsim}
\end{eqnarray}
to which one can get the following three solutions
\begin{eqnarray}
A^{(1)}_{\rm c} &=& 0 \; , \label{eq:Asol10}\\
A^{(2)}_{\rm c} &=& \frac{1}{4} \left(3\cos 2\theta^{}_{13} - \sqrt{1 - 9 \sin^2 2\theta^{}_{13}}\right)\; , \label{eq:Asol20}\\
A^{(3)}_{\rm c} &=& \frac{1}{4} \left(3\cos 2\theta^{}_{13} + \sqrt{1 - 9 \sin^2 2\theta^{}_{13}}\right)\; . \label{eq:Asol30}
\end{eqnarray}
It is evident that the omission of the terms of ${\cal O}(\alpha^{}_{\rm c})$ oversimplifies the cubic equation of $A^{}_{\rm c}$, leading to the incorrect solution $A^{(1)}_{\rm c} = 0$, although we expect a local maximum at a very small value of $A^{}_{\rm c}$ as shown in the left panel of Fig.~\ref{fig:jarlskog}.

\begin{figure}[t!]
	\centering
	\includegraphics[width=0.6\textwidth]{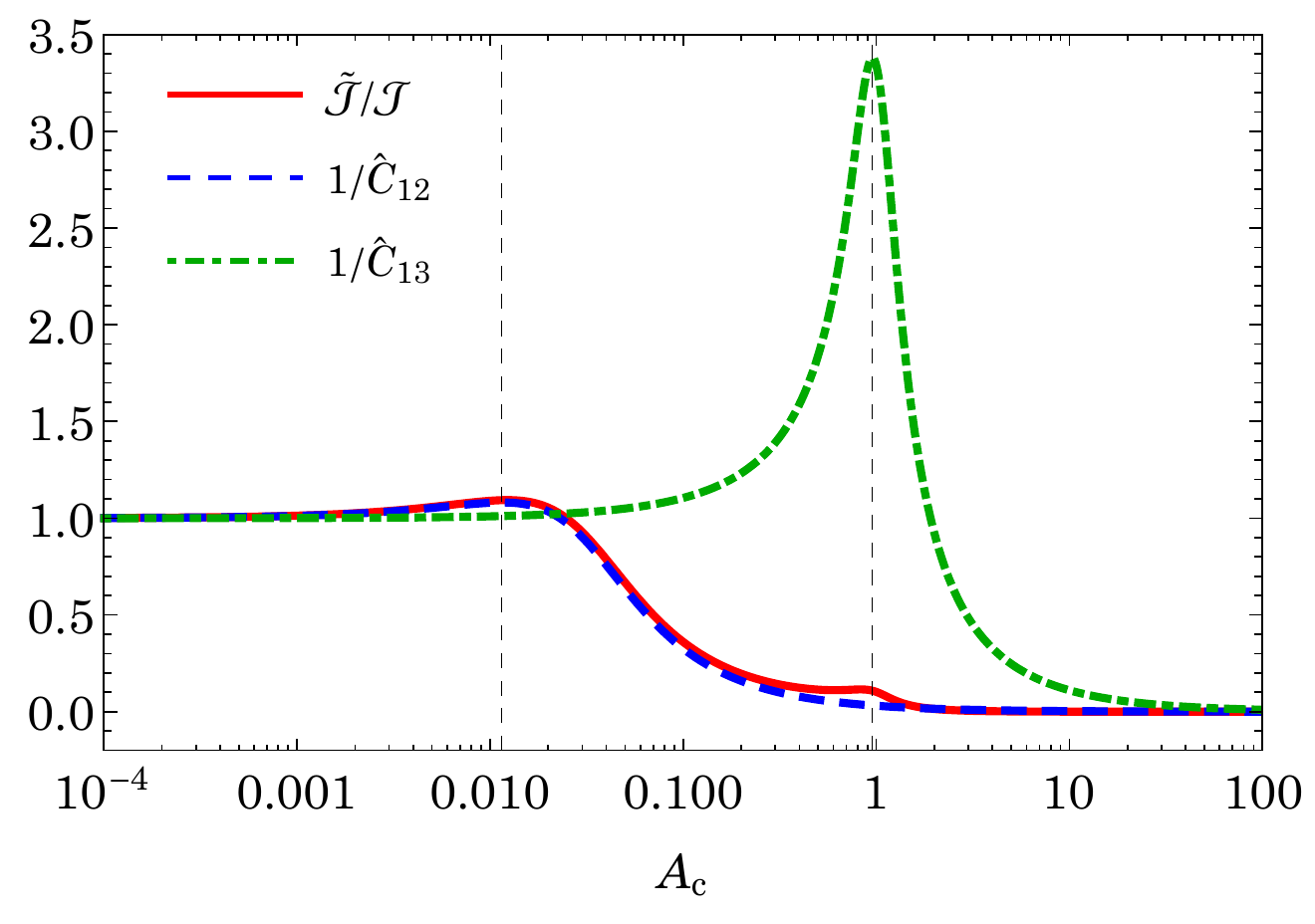}
	\vspace{-0.0cm}
	\caption{The evolution of $\widetilde{\cal J}/{\cal J}$ (the red solid curve), $1/\widehat{C}^{}_{12}$ (the blue dashed curve) and $1/\widehat{C}^{}_{13}$ (the green dot-dashed curve) against the matter parameter $A^{}_{\rm c}$,  where two vertical dashed lines denote respectively the resonance at $\widehat{A}^{}_* = \cos 2\theta^{}_{12}$ and $A^{}_{\rm c} = \cos 2\theta^{}_{13}$. These resonances essentially determine two local maxima of $\widetilde{\cal J}/{\cal J}$.}
	\label{fig:C12C13} 
\end{figure}

\item Retaining all the terms of ${\cal O}(\alpha^{}_{\rm c})$ in Eq.~(\ref{eq:cubic}), we can find the corrected solutions
\begin{eqnarray}
A^{(1)}_{\rm c} &=& \frac{\cos 2\theta^{}_{12}}{\cos^2 \theta^{}_{13}} \alpha^{}_{\rm c} \; , \label{eq:Asol11}\\
A^{(2)}_{\rm c} &=& \frac{1}{4} \left(3\cos 2\theta^{}_{13} + \frac{\cos 2\theta^{}_{12}}{\cos^2 \theta^{}_{13}} \alpha^{}_{\rm c} - \sqrt{1 - 9 \sin^2 2\theta^{}_{13} - \frac{2 \cos 2\theta^{}_{12} \cos 2\theta^{}_{13}}{\cos^2 \theta^{}_{13}} \alpha^{}_{\rm c}}\right)\; , \label{eq:Asol21}\\
A^{(3)}_{\rm c} &=& \frac{1}{4} \left(3\cos 2\theta^{}_{13} + \frac{\cos 2\theta^{}_{12}}{\cos^2 \theta^{}_{13}} \alpha^{}_{\rm c} + \sqrt{1 - 9 \sin^2 2\theta^{}_{13} - \frac{2 \cos 2\theta^{}_{12} \cos 2\theta^{}_{13}}{\cos^2 \theta^{}_{13}} \alpha^{}_{\rm c}}\right)\; , \label{eq:Asol31}
\end{eqnarray}
where the higher-order terms of ${\cal O}(\alpha^2_{\rm c})$ have been omitted. By checking whether the values of the second derivative of $\widetilde{\cal J}/{\cal J}$ with respect to $A^{}_{\rm c}$ at these points are positive or negative, we can identify that $A^{(1)}_{\rm c}$, $A^{(2)}_{\rm c}$ and $A^{(3)}_{\rm c}$ correspond to the first local maximum, the local minimum and the second local maximum of $\widetilde{\cal J}/{\cal J}$, respectively.
\end{itemize}

Substituting Eqs. (\ref{eq:Asol11})-(\ref{eq:Asol31}) into Eq.~(\ref{eq:jexp}), we can find the corresponding three extrema of $\widetilde{\cal J}/{\cal J}$ as below
\begin{eqnarray}
\left.\left(\frac{\widetilde{\cal J}}{\cal J}\right)\right|^{\rm max}_{(1)} &=& \frac{1}{\sin 2\theta^{}_{12}} \left( 1 + \cos 2 \theta^{}_{12} \cos 2\theta^{}_{13} \sec^2 \theta^{}_{13} \alpha^{}_{\rm c} \right) \; , \label{eq:jarmax1} \\
\left.\left(\frac{\widetilde{\cal J}}{\cal J}\right)\right|^{\rm min}_{(2)} &=& \frac{4\sqrt{2} \sec^2_{} \theta^{}_{13} \alpha^{}_{\rm c} }{\sqrt{4 - 3(1 - 3\sin^2 2\theta^{}_{13})^2 + \cos 2\theta^{}_{13}(1 - 9\sin^2 2\theta^{}_{13})^{3/2}}} \; ,  \label{eq:jarmin} \\
\left.\left(\frac{\widetilde{\cal J}}{\cal J}\right)\right|^{\rm max}_{(3)} &=& \frac{4\sqrt{2} \sec^2_{} \theta^{}_{13} \alpha^{}_{\rm c} }{\sqrt{4 - 3(1 - 3\sin^2 2\theta^{}_{13})^2 - \cos 2\theta^{}_{13}(1 - 9\sin^2 2\theta^{}_{13})^{3/2}}} \; ,  \label{eq:jarmax2}
\end{eqnarray}
where only the terms up to ${\cal O}(\alpha^{}_{\rm c})$ are kept. Three extrema of $\widetilde{\cal J}/{\cal J}$, as one can observe from the left panel of Fig.~\ref{fig:jarlskog}, can then be well understood with the help of Eqs.~(\ref{eq:jarmax1})-(\ref{eq:jarmax2}). Some discussions about these approximate analytical results are in order.
\begin{itemize}
\item The first local maximum $(\widetilde{\cal J}/{\cal J})|^{\rm max}_{(1)}$ is approximately $1/\sin 2\theta^{}_{12}$, which is mainly determined by $\widehat{C}^{-1}_{12}|^{}_{A^{}_{\rm c}=A^{(1)}_{\rm c}}$. The other factor $\widehat{C}^{-1}_{13}|^{}_{A^{}_{\rm c}=A^{(1)}_{\rm c}}$ makes a contribution of order ${\cal O}(\alpha^{}_{\rm c})$, as indicated by the second term in the parentheses on the right-hand side of Eq.~(\ref{eq:jarmax1}). To compare the analytical results in Eqs.~(\ref{eq:Asol11}) and (\ref{eq:jarmax1}) with the exact numerical results, we assume a constant matter density $\rho = 3~{\rm g}\cdot {\rm cm}^{-3}$ and an electron number fraction $Y^{}_e = 0.5$, as for the Earth crust. In this case, we have
\begin{eqnarray}
A^{}_{\rm c} = 9.1\times 10^{-2} \cdot \left(\frac{E}{\rm GeV}\right)\cdot \left(\frac{2.5\times 10^{-3}~{\rm eV}^2}{\Delta^{}_{\rm c}}\right) \label{eq:Ac} \; .
\end{eqnarray}
Therefore, one can find out the neutrino energy $E^{(1)}$ for which the first maximum $(\widetilde{\cal J}/{\cal J})|^{\rm max}_{(1)}$ is reached at $A^{}_{\rm c} = A^{(1)}_{\rm c}$, namely,
\begin{eqnarray}
\underline{\rm Analytical}: &~& E^{(1)} = 0.126~{\rm GeV} \; , \quad A^{(1)}_{\rm c} = 0.0115 \; , \quad (\widetilde{\cal J}/{\cal J})|^{\rm max}_{(1)} = 1.09 \; ; \label{eq:amax1}\\
\underline{\rm Numerical}: &~& E^{(1)} = 0.134~{\rm GeV} \; , \quad A^{(1)}_{\rm c} = 0.0122 \; , \quad (\widetilde{\cal J}/{\cal J})|^{\rm max}_{(1)} = 1.09 \; ,
\label{eq:nmax1}
\end{eqnarray}
where the best-fit values of neutrino oscillation parameters have been used in the evaluation of $A^{}_{\rm c}$ in Eq.~(\ref{eq:Ac}). A very good agreement between analytical and numerical results for $A^{(1)}_{\rm c}$ and $(\widetilde{\cal J}/{\cal J})|^{\rm max}_{(1)}$ can be observed from Eqs.~(\ref{eq:amax1}) and (\ref{eq:nmax1}). Looking carefully at Eq.~(\ref{eq:Asol11}), one should be able to recognize that the first local maximum of $\widetilde{\cal J}/{\cal J}$ is indeed achieved when the resonance condition $\widehat{A}^{}_* = A^{}_{\rm c} \cos^2\theta^{}_{13}/\alpha^{}_{\rm c} = \cos 2\theta^{}_{12}$, or equivalently $a\cos^2\theta^{}_{13} = \Delta^{}_{21} \cos 2\theta^{}_{12}$, is satisfied. Therefore, this maximum can be clearly understood as the resonance effect due to the two-flavor neutrino oscillations driven by the mass-squared difference $\Delta^{}_{21}$ and the mixing angle $\theta^{}_{12}$ with the effective matter parameter $a\cos^2\theta^{}_{13}$. The key point is that the matter effects on the effective mixing angle $\widetilde{\theta}^{}_{13}$ for such a small value of $A^{}_{\rm c} = A^{(1)}_{\rm c}$ are basically negligible.

\item The local minimum $(\widetilde{\cal J}/{\cal J})|^{\rm min}_{(2)}$ appears at $A^{}_{\rm c} = A^{(2)}_{\rm c}$. If we assume the same matter density $\rho = 3~{\rm g}\cdot {\rm cm}^{-3}$ and electron number fraction $Y^{}_e = 0.5$, then $A^{(2)}_{\rm c}$ and $(\widetilde{\cal J}/{\cal J})|^{\rm min}_{(2)}$ can be evaluated via Eqs.~(\ref{eq:Asol21}) and (\ref{eq:jarmin}), respectively. For comparison, we present the approximate analytical results and confront it with the exact numerical ones, i.e.,
\begin{eqnarray}
\underline{\rm Analytical}: &~& E^{(2)} = 6.71~{\rm GeV} \; , \quad A^{(2)}_{\rm c} = 0.611 \; , \quad (\widetilde{\cal J}/{\cal J})|^{\rm min}_{(2)} = 0.109 \; ; \label{eq:amax2}\\
\underline{\rm Numerical}: &~& E^{(2)} = 6.69~{\rm GeV} \; , \quad A^{(2)}_{\rm c} = 0.609 \; , \quad (\widetilde{\cal J}/{\cal J})|^{\rm min}_{(2)} = 0.111 \; ,
\label{eq:nmax2}
\end{eqnarray}
where the corresponding values of neutrino energy $E^{(2)}$ for the minimum are also given. To further simplify the analytical results, we expand both Eqs.~(\ref{eq:Asol21}) and (\ref{eq:jarmin}) in terms of $\sin^2 2 \theta^{}_{13}$ and obtain
\begin{eqnarray}
A^{(2)}_{\rm c} &=& \frac{1}{4} \left(2 + 3 \sin^2 2\theta^{}_{13} + 2 \cos 2\theta^{}_{12} \alpha^{}_{\rm c}\right) \approx 0.571\; , \label{eq:aminsim} \\
\left.\left(\frac{\widetilde{\cal J}}{{\cal J}}\right)\right|^{\rm min}_{(2)} &=& \frac{\alpha^{}_{\rm c}}{\cos^2 \theta^{}_{13}} \cdot \frac{4}{\sqrt{1 + 2 \sin^2 2\theta^{}_{13}}} \approx 0.109 \; , \label{eq:jminsim}
\end{eqnarray}
which are well consistent with the original analytical results. In particular, one can see that the local minimum $(\widetilde{\cal J}/{\cal J})|^{\rm max}_{(2)}$ is suppressed by the factor $\alpha^{}_{\rm c}/\cos^2\theta^{}_{13}$, arising from $\widehat{C}^{-1}_{12}|^{}_{A^{}_{\rm c} = A^{(2)}_{\rm c}} \approx \widehat{A}^{-1}_*|^{}_{A^{}_{\rm c} = A^{(2)}_{\rm c}} = \alpha^{}_{\rm c}/(A^{(2)}_{\rm c} \cos^2\theta^{}_{13})$ in the limit of $\widehat{A}^{}_* \gg \cos 2\theta^{}_{12}$.

\item Finally, the second local maximum $(\widetilde{\cal J}/{\cal J})|^{\rm max}_{(3)}$ is reached for $A^{}_{\rm c} = A^{(3)}_{\rm c}$. For the same input parameters, one can get
\begin{eqnarray}
\underline{\rm Analytical}: &~& E^{(3)} = 9.10~{\rm GeV} \; , \quad A^{(3)}_{\rm c} = 0.828 \; , \quad (\widetilde{\cal J}/{\cal J})|^{\rm max}_{(3)} = 0.113 \; ; \label{eq:amax3}\\
\underline{\rm Numerical}: &~& E^{(3)} = 9.10~{\rm GeV} \; , \quad A^{(3)}_{\rm c} = 0.828 \; , \quad (\widetilde{\cal J}/{\cal J})|^{\rm max}_{(3)} = 0.115 \; ,
\label{eq:nmax3}
\end{eqnarray}
where the analytical results are quite accurate as well. Expanding Eqs.~(\ref{eq:Asol31}) and (\ref{eq:jarmax2}) in terms of $\sin^2 2\theta^{}_{13}$, we arrive at
\begin{eqnarray}
A^{(3)}_{\rm c} &=& \frac{1}{2} \left(2 - 3 \sin^2 2\theta^{}_{13} + \cos 2\theta^{}_{12} \tan^2 \theta^{}_{13} \alpha^{}_{\rm c}\right) \approx 0.869\; , \label{eq:amax2sim} \\
\left.\left(\frac{\widetilde{\cal J}}{{\cal J}}\right)\right|^{\rm max}_{(3)} &=& \frac{\alpha^{}_{\rm c}}{\cos^2 \theta^{}_{13}} \cdot \frac{1}{\sin 2\theta^{}_{13}} \approx 0.102 \; , \label{eq:jmax2sim}
\end{eqnarray}
where significant deviations from both the analytical and numerical results in Eqs.~(\ref{eq:amax3}) and (\ref{eq:nmax3}) should be noticed. This can be understood by observing the subtle cancellation in the denominator on the right-hand side of Eq.~(\ref{eq:jarmax2}). However, it becomes clear from Eq.~(\ref{eq:jmax2sim}) that the first factor $\alpha^{}_{\rm c}/\cos^2 \theta^{}_{13}$ on the right-hand side stems from $\widehat{C}^{-1}_{12}$ in the limit of $\widehat{A}^{}_* \gg \cos 2\theta^{}_{12}$ while the second factor $1/\sin 2\theta^{}_{13}$ comes from $\widehat{C}^{-1}_{13}$ under the resonance condition $A^{}_{\rm c} = \cos 2\theta^{}_{13}$.
\end{itemize}

It is worthwhile to mention that the extrema of the effective Jarlskog invariant $\widetilde{\cal J}$ have been previously studied in Ref.~\cite{Yokomakura:2000sv}. In that work,  $\widetilde{\cal J}/{\cal J}$ was first expressed as the ratio of the product of three neutrino mass-squared differences in vacuum to its counterpart in matter, as indicated by the Naumov relation in Eq.~(\ref{eq:naumov}). Then $\widetilde{\cal J}^{-2}$ was written as a quartic function of the matter parameter $a$ in the approximation of $\Delta^{}_{21} \ll \Delta^{}_{31}$. Finally the maxima of $\widetilde{\cal J}$ were found out as the minima of $\widetilde{\cal J}^{-2}$ by solving the cubic equation ${\rm d}(\widetilde{\cal J}^{-2})/{\rm d}a = 0$. It is helpful to make a comparison between the results in Ref.~\cite{Yokomakura:2000sv} and those in the present work. First, starting with the analytical expression of $\widetilde{\cal J}$ in Eq.~(\ref{eq:jexp}), one can immediately verify that $\widetilde{\cal J}^{-2} \approx \widehat{C}^2_{12} \widehat{C}^2_{13}/{\cal J}^2$ is indeed a quartic function of $a$ (or equivalently $A^{}_{\rm c} \equiv a/\Delta^{}_{\rm c}$). Unlike Ref.~\cite{Yokomakura:2000sv}, we can directly find the maxima of $\widetilde{\cal J}/{\cal J}$ by solving the cubic equation derived from ${\rm d}(\widetilde{\cal J}/{\cal J})/{\rm d}A^{}_{\rm c} = 0$. Although the algebraic solutions to the cubic equation are quite standard, the analytical expression $\widetilde{\cal J}/{\cal J} \approx 1/(\widehat{C}^{}_{12}\widehat{C}^{}_{13})$ itself is physically meaningful in the sense that two factors $\widehat{C}^{}_{12}$ and $\widehat{C}^{}_{13}$ are related to two distinct resonances in the two-flavor approximation. Second, with the help of the simple and compact form of $\widetilde{\cal J}$ in Eq.~(\ref{eq:jexp}), it is straightfoward for us to take into account the subleading contributions to the extrema of $\widetilde{\cal J}/{\cal J}$ and the corresponding values of $A^{}_{\rm c}$. Therefore, our results turn out to be more accurate than those in Ref.~\cite{Yokomakura:2000sv}, as one can observe from the previous close comparison between the approximate ananlytical result and the exact numerical one. Third, as the picture of three-flavor neutrino oscillations is now complete and neutrino oscillation parameters are precisely measured, it is timely and necessary to update our understanding of the extrema of the effective Jarlskog invariant by using the latest neutrino oscillation data.

At the end of this section, we shall give some brief comments on the results for antineutrino oscillations and for the IO case. For antineutrino oscillations in the NO case, we can immediately recognize that every term in Eq.~(\ref{eq:cubic}) is negative after making the replacement $A^{}_{\rm c} \rightarrow -A^{}_{\rm c}$. Hence there are no extrema at all for $\widetilde{\cal J}/{\cal J}$. For neutrino oscillations in the IO case, we define $A^{}_{\rm c} \equiv a/\Delta^{}_{\rm c} < 0$ and $\alpha^{}_{\rm c} \equiv \Delta^{}_{21}/\Delta^{}_{\rm c} < 0$. With such definitions, we can derive a cubic equation of $A^{}_{\rm c}$ after calculating the first derivative of $\widetilde{\cal J}/{\cal J}$ and requiring it to be vanishing, which turns out to be exactly Eq.~(\ref{eq:cubic}). Nevertheless, there is only one negative solution, which is $A^{(1)}_{\rm c}$ in Eq. (\ref{eq:Asol11}). This solution corresponds to the unique local maximum of $\widetilde{\cal J}/{\cal J}$ for neutrino oscillations in the IO case. The other two solutions $A^{(2)}_{\rm c}$ in Eq.~(\ref{eq:Asol21}) and $A^{(3)}_{\rm c}$ in Eq.~(\ref{eq:Asol31}) are both positive, which actually give rise to the local minimum and maximum $\widetilde{\cal J}/{\cal J}$ for antineutrino oscillations in the IO case. Following the same procedure as for neutrino oscillations in the NO case, one can easily calculate the extrema of $\widetilde{\cal J}/{\cal J}$. 

\section{Conclusions}\label{sec:conc}

The primary goal of this paper is to improve previous analytical solutions to the RGEs of effective neutrino oscillation parameters, which are first obtained in~\cite{Wang:2019yfp}. We find that with only a simple replacement of $A^{}_{\ast} \equiv a /\Delta^{}_{21}$ by $\widehat{A}^{}_{\ast} \equiv a \cos^{2}_{}\theta^{}_{13}/\Delta^{}_{21}$, the accuracy of the analytical results can be increased remarkably, especially for the results of the effective mixing angle $\widetilde{\theta}^{}_{12}$, the effective Jarlskog invariant $\widetilde{\mathcal J}$ and three effective neutrino mass-squared differences $\widetilde{\Delta}^{}_{ij}$ (for $ij = 21, 31, 32$) in matter. In fact, such a replacement has been noticed in the treatment of matter effects on the flavor conversions of solar neutrinos in the two-flavor approximation. This is in accordance with the previous observations that neutrino oscillations in matter can be efficiently described by two decoupled oscillation modes~\cite{Wang:2019yfp, Denton:2016wmg, Ioannisian:2018qwl, Xing:2019owb}. One is driven by the vacuum oscillation parameters $\{\theta^{}_{12}, \Delta^{}_{21}\}$ with the effective matter parameter $a\cos^2\theta^{}_{13}$, while the other is governed by $\{\theta^{}_{13}, \Delta^{}_{\rm c}\}$ with the ordinary parameter $a$.

Moreover, we have demonstrated that the simple but useful formula $\widetilde{\cal J}/{\cal J} = 1/(\widehat{C}^{}_{12} \widehat{C}^{}_{13})$ with $\widehat{C}^{}_{12} \equiv \sqrt{1 - 2 \widehat{A}^{}_* \cos 2\theta^{}_{12} + \widehat{A}^2_*}$ and $\widehat{C}^{}_{13} \equiv \sqrt{1 - 2 A^{}_{\rm c} \cos 2\theta^{}_{13} + A^2_{\rm c}}$ can be implemented to investigate the basic properties of the matter-corrected Jarlskog invariant $\widetilde{\cal J}$. As an example, we explain the existence and location of two local maxima and one local minimum of $\widetilde{\cal J}/{\cal J} $ for neutrino oscillations in the NO case. The approximate analytical expressions of these extrema and those of the corresponding parameters $A^{}_{\rm c}$ are presented, which are found to be in excellent agreement with the exact numerical results. 

Although it is always possible to compute exactly all the effective neutrino oscillation parameters in a numerical way, the simple and compact analytical results will be helpful in understanding how the matter effects affect neutrino oscillation behaviors in matter. It should be also interesting to see whether the analytical formulas in the present work can be directly used to examine the leptonic unitarity triangles and the probabilities for neutrino oscillations in matter. We hope to come back to these issues in the future works.

\vspace{0.8cm}
\noindent {\Large \bf Acknowledgements}
\vspace{0.3cm}

{\sl The authors thank Prof. Zhi-zhong Xing for helpful discussions. This work was supported in part by the National Natural Science Foundation of China under grant No.~11775232 and No.~11835013, and by the CAS Center for Excellence in Particle Physics.}

\end{document}